\documentclass{article}
\usepackage{spconf,amsmath,graphicx,hyperref}
\usepackage{booktabs}
\usepackage{balance}

\title{Zimtohrli: An Efficient Psychoacoustic Audio Similarity Metric}
\twoauthors
 {Jyrki Alakuijala$^*$, Martin Bruse$^*$\thanks{$^*$ Equal contributions}, Sami Boukortt}
	{Google Research\\
    Zurich, Switzerland}
 {Jozef Marus Coldenhoff, Milos Cernak}
	{Logitech Audio\\
	Lausanne, Switzerland}

\begin{document}
\ninept
\maketitle
\begin{abstract}
This paper introduces Zimtohrli, a novel, full-reference audio similarity metric designed for efficient and perceptually accurate quality assessment. In an era dominated by computationally intensive deep learning models and proprietary legacy standards, there is a pressing need for an interpretable, psychoacoustically-grounded metric that balances performance with practicality. Zimtohrli addresses this gap by combining a 128-bin gammatone filterbank front-end, which models the frequency resolution of the cochlea, with a unique non-linear resonator model that mimics the human eardrum's response to acoustic stimuli. Similarity is computed by comparing perceptually-mapped spectrograms using modified Dynamic Time Warping (DTW) and Neurogram Similarity Index Measure (NSIM) algorithms, which incorporate novel non-linearities to better align with human judgment.
Zimtohrli achieves superior performance to the baseline open-source ViSQOL metric, and significantly narrows the performance gap with the latest commercial POLQA metric. It offers a compelling balance of perceptual relevance and computational efficiency, positioning it as a strong alternative for modern audio engineering applications, from codec development to the evaluation of generative audio systems.
\end{abstract}

\section{Introduction}
\label{sec:intro}

A quest for high-fidelity audio, from real-time telecommunications to sophisticated generative modeling, is fundamentally constrained by our ability to measure perceptual quality. Objective, full-reference (FR) metrics, which compute a quality score by comparing a processed or degraded signal to its original, are indispensable tools for algorithm development, codec evaluation, and automated quality assurance. They provide a scalable and cost-effective alternative to laborious subjective listening tests.

However, the very nature of audio impairments has evolved dramatically, creating a moving target for metric design. Early metrics were developed to quantify the relatively predictable distortions introduced by narrowband speech codecs and linear filtering in traditional telephone networks. The technological landscape has since shifted. Modern systems, particularly those based on Voice over IP (VoIP), introduce a host of complex, non-linear temporal artifacts such as time-warping, jitter, and glitches from packet loss concealment, which proved challenging for older models. More recently, the proliferation of deep neural network (DNN)-based systems for speech enhancement and generative audio synthesis presents an entirely new frontier of challenges. These models synthesize entirely new waveforms that can be perceptually excellent but bear little mathematical resemblance to any single reference signal, confounding traditional similarity-based approaches.

This co-evolution of audio processing and quality assessment has created a continuous development, where new forms of degradation necessitate the development of new metrics. The transition from PESQ~\cite{rix2001perceptual} to POLQA~\cite{beerends2013perceptual} was a direct response to the rise of VoIP; the current shift towards learned metrics is a reaction to the complex, non-linear outputs of deep learning systems. This dynamic reveals that the state of the art in measurement often lags the state of the art in processing. It also presents a trilemma for researchers and practitioners in selecting an appropriate metric:

\textbf{Classical Standards (e.g., POLQA):} These metrics are robust, well-validated, and widely adopted as industry benchmarks. However, they are often proprietary, computationally demanding, and their implementations are frequently treated as a "black box," which limits academic exploration, modification, and interpretation.

\textbf{Open-Source Signal-Processing Metrics (e.g., ViSQOL):} These metrics provide transparent and valuable alternatives. While effective for many VoIP-related degradations, they may not capture all perceptual nuances and can be outperformed by newer approaches in certain domains~\cite{chinen2020visqol}.

\textbf{Learned Deep Learning Metrics (e.g., CDPAM)}: These data-driven models represent the state of the art in correlation with human perception~\cite{manocha2021cdpam}. Yet, they are computationally expensive, require massive, costly-to-collect human-labeled datasets for training, and can suffer from poor generalization when confronted with degradation types not seen during training.

Compounding this trilemma is a foundational crisis emerging within the field: the growing body of evidence that signal similarity is an unreliable and often misleading proxy for perceptual quality. Seminal research has demonstrated that leading FR metrics are highly sensitive to perceptually irrelevant factors, such as the specific acoustic environment of the reference recording, while being insensitive to genuine quality differences that are obvious to human listeners~\cite{manocha22interspeech}. This challenges the validity of the entire FR paradigm and suggests a need to redefine its objective. The goal should perhaps not be to measure mathematical similarity, but rather to model the human auditory system's remarkable invariance to certain types of dissimilarity. Humans effortlessly disentangle speech content from acoustic characteristics like speaker timbre or room reverberation, a feature that lacks in current metrics.

This paper introduces Zimtohrli, a novel FR metric designed to navigate this complex landscape. It is a psychoacoustically-grounded, signal-processing-based approach that is transparent, computationally efficient, and does not require large-scale supervised training. By incorporating a novel non-linear resonator model and modified comparison algorithms, Zimtohrli aims to model human auditory perception more faithfully than traditional linear methods. It represents a strategic effort to create a more future-proof metric by relying on fundamental psychoacoustic principles rather than tuning to specific, transient artifact types, thereby providing a robust and practical tool for the next generation of audio quality assessment.

\section{Related work}
\label{related}
In response to the need for automated assessment, the International Telecommunication Union (ITU-T) standardized a series of influential FR metrics based on computational models of psychoacoustics. These models attempt to predict subjective scores by simulating key aspects of the human auditory pathway.

For many years, the PESQ (ITU-T P.862) was the workhorse of the telecommunications industry. The PESQ algorithm aligns the degraded signal with the clean reference, and transforms both signals into an internal psychoacoustic representation.
The perceptual difference between the reference and degraded representations is then calculated and mapped to a MOS-like score. However, PESQ was designed and calibrated for the types of degradation common in the circuit-switched telephone networks of its era. As a result, its underlying model proved poor when faced with the more complex impairments of modern packet-switched networks.
The POLQA method (ITU-T P.863) was developed as the direct successor to PESQ, engineered specifically to address these shortcomings and meet the demands of modern communication systems. It performs robust temporal alignment~\cite{beerends2013perceptual}, expanded bandwidth support up to full-band audio, uses improved psychoacoustic model to better account for the effects of background noise, reverberation, and the non-linear processing artifacts introduced by modern voice enhancement algorithms.

Distinct from the ITU-T lineage, the Virtual Speech Quality Objective Listener (ViSQOL) was developed as a robust, open-source FR metric that has gained significant traction in both academia and industry.
ViSQOL's methodology is fundamentally different from that of PESQ and POLQA, as it operates by directly comparing time-frequency representations of the audio signals. The core algorithm consists of four main steps: (i) Both the reference and degraded audio signals are first transformed into spectrograms using a gammatone filterbank.
(ii) The reference spectrogram is then segmented into a series of overlapping time-frequency "patches" for localized comparison. (iii) The central comparison is performed using the Neurogram Similarity Index Measure (NSIM) that is an auditory-domain adaptation of the highly successful Structural Similarity Index Measure (SSIM) from the field of image quality assessment.
For each patch from the reference spectrogram, NSIM is used to search for and score the best-matching patch within a corresponding temporal window in the degraded spectrogram. This patch-based search provides inherent robustness to small temporal misalignments. (vi) The average NSIM similarity score across all patches is computed and then mapped to a final MOS score. This mapping is typically performed by a machine learning model, such as a support vector regression (SVR) or, in newer versions, a deep lattice network models.

The last decade has witnessed a paradigm shift in metric design, moving away from models engineered based on explicit psychoacoustic principles and towards models that are learned directly from vast amounts of data.
This reflects a fundamental divergence in design philosophy, from a "simulationist" approach that attempts to build an explicit computational model of the human auditory pathway, to a "behaviorist" approach that learns a function to mimic the  output of human judgment without necessarily modeling the internal mechanism. The Deep Perceptual Audio Metric (DPAM)~\cite{manocha20_interspeech} and CDPAM (Contrastive DPAM)~\cite{manocha2021cdpam} pioneered this behavioral approach.
Frameworks such as NORESQA~\cite{manocha2021noresqa} and NOMAD~\cite{ragano2024nomad} were trained to assess the quality of a degraded signal relative to any clean reference signal, not just its content-matched original.
The CoRN framework~\cite{manocha2024corn} proposes the joint training of full and no reference models that share a common feature encoder.

\section{Zimtohrli}
\label{zimtohrli}
Zimtohlri is a novel full-band full-reference metric grounded in psychoacoustic principles, designed to be both perceptually relevant and computationally efficient. It can be understood as an evolution of the spectrogram-similarity paradigm, adopting a successful foundation but integrating novel, physics-inspired non-linear components to better model the complexities of the human auditory system. The metric consists of three main stages: a psychoacoustic front-end, a perceptual spectrogram generation stage, and a modified spectrogram comparison stage. For all processing, incoming audio signals are resampled to a 48 kHz sampling rate. It is implemented here\footnote{\url{https://github.com/google/zimtohrli}}.

\subsection{Psychoacoustic model}

The front-end of Zimtohrli transforms the raw audio waveform into a representation that approximates the initial processing performed by the human ear. It features two parallel processing paths.

The primary path mimics the frequency analysis of the human cochlea using a 128-bin, 3rd-order complex gammatone filterbank. This type of filterbank is well-established in auditory modeling. The center frequencies of the 128 filters are distributed along the spectrum not linearly, but in equal steps of Equivalent Rectangular Bandwidth (ERB), following the perceptual scales derived from the notched-noise data of Glasberg \& Moore (1990)~\cite{glasberg1990derivation}. A key feature of the implementation is that the bandwidth of each filter is dynamically adjusted based on its neighboring filters. This adaptive bandwidth then directly modifies the filter's integration coefficient $C$, according to the empirically derived formula: $C\approx0.9996^{(BW_{Hz}*0.7323)}$, where $BW_{Hz}$ is the filter's bandwidth in Hertz.

A central innovation in Zimtohrli is the inclusion of a parallel resonator path designed to introduce plausible, bio-inspired non-linearity into the processing pipeline. This path is intended to model the non-linear response of mass-spring systems within the ear, most notably the tympanic membrane (eardrum). The input signal is passed through a 32-bin time-based linear filter to feed this resonator model, which performs a complex spectral shifting of energy. It is hypothesized that this non-linear component allows the metric to capture perceptual phenomena related to transient response and mechanical damping that purely linear filterbank models may miss.

\subsection{Perceptual Spectrogram Generation}

The outputs of the front-end are used to generate a perceptually-mapped spectrogram. The raw energy values from the gammatone filterbank are first subsampled to the target 85 Hz frame rate. A Sigmoid function is then applied to spread energy values between adjacent frequency bins. This operation is designed to model the overlapping nature of auditory nerve excitation patterns, where a single tone excites a range of nerve fibers centered around the characteristic frequency.

The subsampled and spread representation is then passed to the \texttt{LoudnessDb} function, which computes a single time-frame (a column vector) of the final spectrogram. This function performs two critical psychoacoustic transformations, noise floor addition and logarithmic transformation. In the former one, a small bias, representing a perceptual noise floor, is added to all energy values before any logarithmic compression. This serves two purposes: it stabilizes the subsequent logarithmic transform for near-zero energy values, preventing numerical instability, and it ensures that signal components with energy below this floor are rendered perceptually insignificant in the final representation. A logarithmic transform is applied to the energy-plus-bias values. This well-known operation compresses the dynamic range of the signal in a way that approximates the human ear's perception of loudness. The transformation is scaled by an array of frequency-dependent multipliers to account for the ear's varying sensitivity across the spectrum.

\subsection{Modified Spectrogram Comparison}

The final stage of the metric involves comparing the perceptual spectrogram of the degraded signal to that of the reference signal. This stage incorporates several modifications to standard comparison algorithms to better align with human similarity judgments.

To account for overall loudness differences—which can be a quality factor but can also confound similarity measures—the spectrograms of the reference and degraded signals are partially normalized. The maximum values of both spectrograms are measured, and the signal levels are adjusted to bring these maximums 82\% closer to each other. This partial normalization reduces the impact of simple gain differences without completely eliminating them.

A modified DTW algorithm is employed for the fine-grained temporal alignment of the two spectrograms. In a standard DTW, the cost of warping is typically linear. In Zimtohrli's modification, the delta (cost) values computed during the warping path search are raised to a power less than 1. This non-linear transformation of the cost function creates a "soft" preference for no time warping, penalizing excessive temporal distortion more gracefully and in a more perceptually plausible manner than a standard linear-cost DTW.

After alignment, a modified version of the NSIM is used to compute the final similarity score. The specific modifications involve the introduction of new non-linearities into the core of the NSIM calculation. These are designed to capture more complex perceptual relationships between the spectrogram patches than the standard NSIM formulation allows.

The model's parameters were tuned by alternating stochastic optimization (using a $\beta(1/5, 1)$ distribution for perturbations with amplitude scaling control) and the downhill simplex method. This optimization process utilized roughly 200,000 CPU hours. The model was trained to maximize the Spearman correlation (Eq.~(\ref{eq:rho})) between its similarity scores and the human-rated scores sourced from ten distinct audio similarity corpora including various kinds of audio samples from speech to music. The following datasets with existing human-rated scores were used: CORESVNET\footnote{\url{https://listening-test.coresv.net/}}, ODAQ~\cite{torcoli2024odaq}, PerceptualAudio\footnote{\url{https://github.com/pranaymanocha/PerceptualAudio}}, PEASS\_DB\footnote{\url{https://gitlab.inria.fr/bass-db/peass}}, SEBASS-DB~\cite{kastner2022sebass}, TCD-VoIP~\cite{harte2015tcd}, and two Google internal proprietary datasets. To counteract the deteriorating impact of the Spearman correlation's strong discretization effect, we represented the exact scores as distributions during optimization.


The Zimtohrli model only compares a single reference channel with a single distortion channel. But all the command line tools have the ability to compute the L2 of the distances between reference and distortion for each channel.

\section{Experiments and results}
\label{experiments}


To validate Zimtohrli's performance, we conducted a comprehensive evaluation using four publicly available, full-band datasets containing human-rated scores. The datasets cover a wide range of degradations:
\begin{itemize}
    \item \textbf{NISQA TEST FOR \& P501}\footnote{\url{https://github.com/gabrielmittag/NISQA/wiki/NISQA-Corpus}}: These datasets contain speech with simulated distortions from various codecs, background noises, packet loss, and clipping, as well as live conditions from communication platforms like Zoom and WhatsApp.
    \item \textbf{EARS-EMO-OpenACE}\footnote{\url{https://huggingface.co/datasets/mcernak/EARS-EMO-OpenACE}}: This dataset contains emotional speech encoded with modern codecs (Opus, EVS, LC3, LC3Plus) at 16 kbps, providing a challenging test case with real-world variability.
    \item \textbf{SQAM-POP-MUSIC}: To assess performance on non-speech content, we used vocal-orchestra and pop-music signals from the SQAM database\footnote{\url{https://tech.ebu.ch/publications/sqamcd}}~\cite{EBU-SQAM}. These were processed by two high-quality codecs: the legacy Opus codec at 128 kbps and the neural Descript Audio Codec~\cite{kumar2023high} at 8 kbps. The codecs creates a contrastive pair: while the former is the legacy codec used for YouTube, high bit rate but low computations, the later is recent neural codec, with low bitrate but high computations. We collected new human ratings for this subset via a MUSHRA listening test~\cite{BS_1534_3}, with the mid-quality (7000 Hz) anchor, and truncated listening stimuli to the first 10 seconds. 
\end{itemize}

Zimtohrli's performance was benchmarked against the proprietary industry standard POLQA V3 (as a reference) and the open-source ViSQOL V3 (as a baseline). 

We report three standard correlation coefficients between objective method scores ($x_i$) and subjective listener ratings ($y_i$) with $n$ as the number of observations: (i) PLCC (Pearson’s linear correlation coefficient), (ii) SRCC (Spearman’s rank correlation coefficient), and (iii) KRCC (Kendall’s rank correlation coefficient). PLCC (Eq.~\ref{eq:r}) measures the strength of a linear relationship of the scores. SRCC, defined by Eq.~\ref{eq:rho}, where $d$ is the difference between the ranks of paired data points and $n$ is the number of data points, evaluates how well the relationship between two variables can be described by a monotonic (not necessarily linear) function by correlating rank orders. KRCC (Eq.~\ref{eq:tau}) measures provide a more robust (pairwise) assessment of ordinal agreement and being less influenced by large local rank shifts than SRCC. Reporting all three gives a fuller view: PLCC for linear predictive accuracy, SRCC for ordering, KRCC for robust ordinal consistency.

\begin{equation}
\label{eq:r}
\text{PLCC} = r = \frac{\sum_{i=1}^n (x_i-\bar{x})(y_i-\bar{y})}{\sqrt{\sum_{i=1}^n (x_i-\bar{x})^2};\sqrt{\sum_{i=1}^n (y_i-\bar{y})^2}}
\end{equation}

\begin{equation}
\label{eq:rho}
\text{SRCC} = \rho = 1 - \frac{6\sum_{i=1}^n d_i^2}{n(n^2-1)}
\end{equation}

\begin{equation}
\label{eq:tau}
\text{KRCC} = \tau = 
\frac{2}{n(n-1)} \sum_{1\le i<j\le n} \operatorname{sgn}!\big((x_i-x_j)(y_i-y_j)\big)     
\end{equation}



\begin{table*}[t]
\centering \caption{Correlation between objective metrics and subjective scores (PLCC ($r$)/ SRCC ($\rho$)/ KRCC ($\tau$)). Higher is better.} \small 
\begin{tabular}{l||ccc||ccc|ccc|ccc}
\toprule
Dataset & \multicolumn{3}{c||}{POLQA V3 (ref)} & \multicolumn{6}{c|}{VISQOL V3 (baseline)} & \multicolumn{3}{c}{Zimtohrli (proposed)} \\
& & & & \multicolumn{3}{c}{\textit{audio mode 48kHz}} & \multicolumn{3}{c|}{\textit{speech mode 16kHz}} & \multicolumn{3}{c}{\textit{always 48kHz}} \\
\cline{5-10}
& $r$ & $\rho$ & $\tau$ & $r$ & $\rho$ & $\tau$ & $r$ & $\rho$ & $\tau$ & $r$ & $\rho$ & $\tau$ \\
\midrule
NISQA\_TEST\_FOR ($n=240$)& 0.85 & 0.85 & 0.66 & 
0.41 & 0.37 & 0.26 & 
0.61 & 0.60 & 0.43 &
\textbf{0.70} & \textbf{0.64} & \textbf{0.46} \\
NISQA\_TEST\_P501 ($n=240$)& 0.88 & 0.86 & 0.68 & 
0.48 & 0.44 & 0.31 & 
\textbf{0.76} & \textbf{0.79} & \textbf{0.59} &
0.74 & 0.70 & 0.52\\ 
EARS-EMO-OpenACE ($n=216)$& 0.65 & 0.66 & 0.47 & 
0.42 & 0.49 & 0.30 & 
0.47 & 0.56 & 0.33 &
\textbf{0.57} & \textbf{0.65} & \textbf{0.44}\\
SQAM-POP-MUSIC ($n=30$) & 
0.88 & 0.78 & 0.59 &
0.81 & 0.83 & 0.66 &
0.80 & 0.76 & 0.56 &
\textbf{0.95} & \textbf{0.94} & \textbf{0.81}\\
\bottomrule
\end{tabular} \label{tab:correlations} \end{table*}

\begin{figure}[th]
    \centering
    \includegraphics[width=0.98\linewidth]{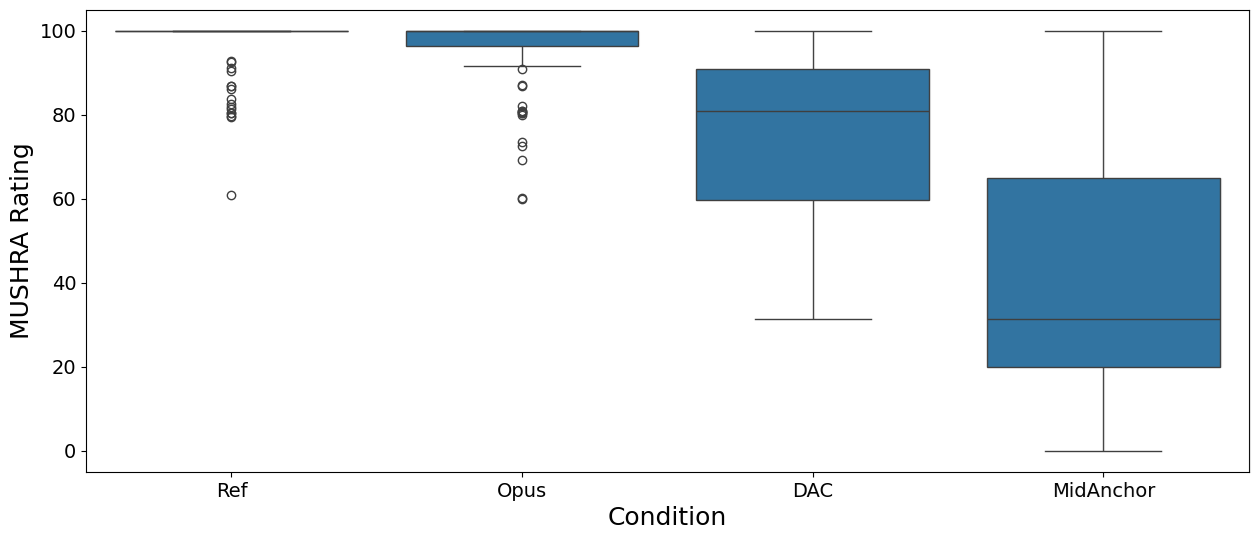}

   \caption{
      \centering
      {Subjective assessment of the SQAM-POP-MUSIC test.
      }}
    \label{fig:SQAM}
\end{figure}

\subsection{Results and discussion}
\label{results}

Fig.~\ref{fig:SQAM} shows performane of Opus 128 kbps and DAC neural coding at 8 kbps obtained using a MUSHRA listening test. 15 listeners rated audio quality of encoded music signals, and the ratings were subsequently used in the correlation analysis. DAC often changes the volume up and down or has other distortions that made it clearly worse encoding.

The correlation results, summarized in Table~\ref{tab:correlations}, demonstrate Zimtohrli's strong performance across diverse conditions. All reported correlations are statistically significant ($p < 0.001$), with the sample sizes ranging from 30 to 240 audio pairs.

As expected, the proprietary POLQA metric\footnote{\url{http://polqa.info/}} demonstrated superior performance across the speech datasets. However, Zimtohrli significantly narrows the performance gap, showing strong correlations ($r > 0.70$) on the NISQA datasets and outperforming the ViSQOL baseline\footnote{\url{https://github.com/google/visqol}}. Average correlations per method are as follows: POLQA V3 0.73, Zimtohrli 0.68, ViSQOL speech 0.60, and ViSQOL audio 0.48.
ViSQOL thus achieves relative performance drop to POLQA about 34\% and 18\% for the audio and speech mode respectively, whereas Zimtohrli has shrunk the performance drop to about 7\%.

Most notably, Zimtohrli achieved a remarkably strong correlation ($r = 0.95$) on the SQAM-POP-MUSIC audio signals, surpassing POLQA and indicating its robustness for general audio content beyond speech. In contrast, the ViSQOL baseline is highly context-dependent, requiring different modes for speech and audio and showing inconsistent performance. The moderate correlation results for all metrics on the EARS-EMO-OpenACE dataset suggest that assessing encoded emotional speech remains a significant challenge for current full-reference methods.

No significant gender bias detected. All metrics perform consistently across male and female speakers, with POLQA showing slightly stronger correlations for male speakers.

In addition to perceptual accuracy, computational efficiency is a key design goal. As shown in Table~\ref{tab:runtime}, Zimtohrli is significantly more efficient than the ViSQOL audio mode baseline running at 48kHz, processing a 10-second audio clip over 5 times faster while using less memory.

\begin{table}
\centering \caption{Runtime performance comparison between Zimtohrli, and ViSQOL in audio mode. Both metrics are run using their Python bindings. Numbers shown are based on 10 seconds of input audio on an Intel i9-9900k. } \small 
\begin{tabular}{l||ccc||ccc||ccc}
\toprule
Metric          & Runtime (10s audio) & Memory (10s audio) \\
\midrule
ViSQOL (48k) & 5.6 s ± 96.8 ms     & 271.41 MiB              \\
Zimtohrli       & 1.01 s ± 11.1 ms    & 250.10 MiB         \\
\bottomrule
\end{tabular} \label{tab:runtime} \end{table}

\section{Conclusion and future work}
\label{conclusions}
This paper has introduced Zimtohrli, a novel full-band, full-reference audio similarity metric designed for universal assessment across both speech and general audio. Zimtohrli is inspired to hear the `smallest` degradations in audio compression systems. Grounded in psychoacoustic principles and enhanced with novel non-linear modeling, Zimtohrli was developed to provide a transparent, efficient, and perceptually relevant open-source alternative to proprietary standards and computationally expensive deep learning models.

Our experimental evaluation demonstrates that Zimtohrli achieves this goal. It significantly narrows the performance gap to the commercial POLQA v3 metric on standard speech quality datasets and consistently outperforms the open-source ViSQOL baseline. A key finding is Zimtohrli's exceptional performance on audio signals, where it surpassed all other tested metrics, highlighting its versatility. Furthermore, it achieves this performance with over 5x greater computational efficiency than ViSQOL, making it a practical tool for large-scale evaluation.

The results also highlight areas for future work. The moderate performance of all metrics on encoded emotional speech indicates that capturing the nuances of such complex signals remains an open challenge. The most compelling future direction, however, is the development of a fully differentiable version of the Zimtohrli pipeline. Such a model could serve as a perceptually-grounded loss function for training the next generation of generative audio models, bridging the gap between traditional signal processing and data-driven optimization. By offering a strong balance of performance, efficiency, and interpretability, Zimtohrli provides a valuable new tool\footnote{\url{https://pypi.org/project/zimtohrli/}} for the audio engineering community.


\vfill\pagebreak

\balance
\bibliographystyle{IEEEbib}
\bibliography{refs}

\end{document}